\documentclass[structabstract]{aa} 
\usepackage{graphicx}
\usepackage{longtable}
\usepackage[a4paper=true,hypertex]{hyperref}
\usepackage{txfonts}
\usepackage{natbib}
\usepackage{epsfig}
\usepackage{longtable}
\usepackage{times}
\usepackage[english]{babel}
\usepackage{url}

\bibpunct{(}{)}{;}{a}{}{,}

\newcommand{\src}{\mbox{IGR\,J16493$-$4348 }}
\newcommand{\sw} {\textit{Swift }}

\newcommand{\chandra}{\textit{Chandra }}
\newcommand {\nh} {N$_{\rm H}$ }

\newcommand {\ferg} {erg cm$^{-2}$ s$^{-1}$}
\newcommand {\cmmdue} {cm$^{-2}$}
\newcommand {\hcm} {\hbox {\ifmmode $ atom cm$^{-2}\else atom cm$^{-2}$\fi}}
\newcommand {\chisq} {$\chi^{2}$}

\newcommand {\msun} {M$_{\odot}$}
\newcommand {\rsun} {R$_{\odot}$}
\begin{document}

\title{Evidence for a resonant cyclotron line in IGR\,J16493$-$4348\\ from the Swift-BAT hard X-ray survey}
\authorrunning{A. D'A\`i et al.}
\titlerunning{A cyclotron line in IGR~J16493$-$4348}

\author{A.D'A\`i$^{1}$, G.\ Cusumano$^{2}$, V.\ La Parola$^{2}$, A.\ Segreto$^{2}$,
T.\ Di Salvo$^{1}$, R.\ Iaria $^{1}$, N.R.\ Robba$^{1}$}
\institute{ 
Dipartimento di Fisica, Universit\`a di Palermo, Via Archirafi 36, I-90123, Palermo, Italy
\email{antonino.dai@unipa.it} 
\and
INAF, Istituto di Astrofisica Spaziale e Fisica Cosmica, Via U.\ La Malfa 153, I-90146 Palermo, Italy}

\date{}

\abstract{}{}{}{}{} 
  \abstract
%context
{Resonant absorption  cyclotron features are a key  diagnostic tool to
  directly  measure the strength  of the  magnetic field  of accreting
  neutron stars.   However, typical values for  cyclotron features lie
  in the high-energy  part of the spectrum between 20  keV and 50 keV,
  where detection  is often damped  by the low statistics  from single
  pointed observations.}
% aims heading (mandatory)
{We show that long-term monitoring campaign performed with Swift-BAT of persistently, 
but faint, accreting high-mass X-ray binaries  is able to reveal in their 
spectra the  presence of  cyclotron features.}
% methods heading (mandatory) 
%%
{We extracted the average Swift-BAT 15-150 keV spectrum from the 54 months 
long Swift-BAT survey of the high-mass X-ray source IGR~J16493$-$4348. 
To constrain the broadband spectrum 
we used soft X-ray spectra from Swift-XRT and Suzaku pointed observations.}
% results heading (mandatory)
{ 
We model the spectra using a set of phenomenological models usually adopted
to describe the energy spectrum of accreting high-mass X-ray binaries; irrespective 
of the models we used, we found significant improvements in the spectral fits adding to
the models a broad (10 keV width) absorption feature, with best-fitting energy estimate between 30 and 33 keV, 
that we interpret as evidence for a  resonant cyclotron absorption feature. 
We also discuss instrumental issues related to the use of Swift-BAT for this kind
of studies and the statistical method to weight the confidence level of this detection.  
  Correcting  for  the  gravitational  redshift of  a  1.4  M$_{\sun}$
  neutron star,  the inferred surface magnetic field is $B_{\rm surf}
  \sim 3.7  \times 10^{12}$  Gauss. The spectral parameters of IGR~J16493$-$4348 fit well 
  with empirical correlations observed when the whole sample of high-mass binaries with detected cyclotron
features is considered.}
%conclusions
{}

\maketitle

%%%%%%%%%%%%%%%%%%%%%%
%%%%%%%%%%%%%%%%%%%%%%
%%%%%%%%%%%%%%%%%%%%%%
%%%%%%%%%%%%%%%%%%%%%%
%%%%%%%%%%%%%%%%%%%%%%
%%%%%%%%%%%%%%%%%%%%%% 

\section{Introduction\label{intro}}
%preamble1
The Burst Alert Telescope  (BAT, \citealp{barthelmy05}) on board Swift
\citep{gehrels04} has  been performing a continuous  monitoring of the
sky in the hard X-ray  energy range (15--150 keV) since November 2004.
The telescope, thanks  to its large field of  view (1.4 steradian half
coded) and its  pointing strategy, covers a fraction  between 50\% and
80\% of the sky every day.   This has allowed the detection of many of
the    new   INTEGRAL    High-Mass   X-ray    Binaries   \citep[HMXBs,
  e.g.][]{cusumano10b}  and the  collection of  their long  term light
curves  and spectra.   The  long and  continuous  monitoring of  these
sources allows  to investigate the intrinsic  emission variability, to
search for  long periodicities (orbital  periods) and to  discover the
presence of  eclipse events.  The role of  Swift-BAT (hereafter simply
BAT) is therefore fundamental to unveil the nature and the geometry of
these binary  systems. Moreover BAT collects  their long-term averaged
energy spectra in the 15.0-150  keV band.  For HMXBs this energy range
is  extremely  important  as  resonant cyclotron  scattering  features
(CRSFs)  are usually  observed  in  this part  of  the X-ray  spectrum
\citep{heindl04}.

In this  paper we analyse  the soft and  hard X-ray data  collected by
BAT,  INTEGRAL and  by pointed  Suzaku and  Swift-XRT  observations on
\src.  This  source was discovered by INTEGRAL  in 2004 \citep{bird04}
and it was initially  associated with the radio pulsar PSR~J1649--4349
because of a spatial coincidence.  A later INTEGRAL observation with a
deep  exposure allowed  to reduce  the positional  uncertainty  and to
reject   the   pulsar   association  \citep{atel457}.    A   follow-up
observation with \chandra found a soft X-ray counterpart at a position
of RA(J2000)  = 16\degr 49\arcmin 26.92\arcsec;  Dec(J2000) = -43\degr
49\arcmin  8.96\arcsec \citep{atel654}  allowing the  association with
the   infrared   counterpart   2MASS~J1642695--4349090,  a   B0.5   Ib
supergiant.   \citet{nespoli10} performed  K-band spectroscopy  of the
inferred counterpart,  confirming the B0.5–1  Ia-Ib companion spectral
type; the infrared extinction, translated into the equivalent hydrogen
column  density,  was  estimated  to  be (2.92  $\pm$  1.96)  $\times$
10$^{22}$ \cmmdue,  a value  that is lower  with respect to  the X-ray
absorption, thus indicating  that part of the X-ray  absorption may be
local to  the compact object.  \citet{cusumano10}  found a periodicity
of  6.732 $\pm$  0.002 d,  interpreted as  the orbital  period  of the
system. The folded  light curve shows presence of  an eclipse, lasting
$\sim$ 12\% of the P$_{orb}$. Assuming a 1.4 \msun\, neutron star (NS)
and 32 \rsun\, for the companion  star, the system has most likely a low
eccentricity value ($e  \leq$ 0.15).  Recently, \citet{atel2766} found
in RXTE observations evidence for a shorter periodicity at 1069 $\pm$
7 s, interpreted as the NS spin  period.

 A spectral analysis using non simultaneous data from Swift-XRT and INTEGRAL
 \citep{hill08} showed that the broadband X-ray spectrum could be well
 fitted  by an  absorbed (\nh=5.6  $\times$ 10$^{22}$  \cmmdue) power-law
 ($\Gamma$ = 0.6$\pm$0.3) with a  high-energy cut-off at $\sim$ 17 keV.
 \citet{morris09}   analysing  Suzaku   data  of   \src,   proposed  an
 alternative modelling of the  X-ray spectrum using a partial covering
 component multiplied by a  simple power-law. The covering fraction of
 the total  emission was estimated  in 0.62$\pm$0.07, while  the local
 value of the \nh was estimated in $\sim$ 30 $\times$ 10$^{22}$ \cmmdue.

\section{Data Reduction\label{data}}

The raw  BAT survey data of the  first 54 months of  the Swift mission
were        retrieved       from       the        HEASARC       public
archive\footnote{\url{http://heasarc.gsfc.nasa.gov/docs/archive.html}}
and  processed  with  a  dedicated  software  \citep{segreto10},  that
performs screening,  mosaicking and source  detection on BAT  data and
produces spectra  and light  curves for any  given sky  position.  The
light curve of \src was extracted in the 15--150 keV energy range with
the maximum  available time resolution  ($\sim 300$ s).   The spectrum
was obtained by  extracting the source count rates  in 16 energy bands
and    analysed     using    the    BAT     spectral    redistribution
matrix\footnote{\url{http://heasarc.gsfc.nasa.gov/docs/heasarc/caldb/data/swift/bat/index.html}}.
The last three energy channels  (100-150 keV range) were rebinned into
a single channel  in order to have a S/N ratio  above 3 $\sigma$.  The
source  is detected  at a  significance  level of  $\sim$ 21  standard
deviations.   The  average  count  rate  in the  BAT  light  curve  is
$1.03\times  10^{-4}$ count  s$^{-1}$ pixel$^{-1}$.   When considering
the light curve at the highest resolution the maximum of the deviation
from the average rate is about $7\sigma$, corresponding to an increase
in  rate of  a factor  $\sim20$.   Although the  count rate  indicates
brightness variations  in the accreting  source, we found  no evidence
for spectral  shape variability, when  spectra collected on  one month
time  span were  fitted using  a simple  power-law model,  with photon
indices  consistent with  the average  value within  the  errors.  The
15--50  keV and  50--150  keV time-averaged  fluxes are  (3.0$\pm$0.3)
$\times$   10$^{-11}$   \ferg   and   (1.4$_{-0.3}^{+0.1}$)   $\times$
10$^{-11}$ \ferg, respectively.

We  analysed also  INTEGRAL-ISGRI (hereafter  simply  ISGRI) long-term
data   obtained  from   the  Integral   high-level   products  archive
HEAVENS\footnote{\url{http://www.isdc.unige.ch/heavens/}}.  ISGRI data
were used in the 15$-$150 keV energy range.

To have the necessary coverage of  the soft X-ray band we used data of
the  Swift-XRT  \citep[hereafter  simply XRT,][]{burrows05}  observation
performed on 2006 March 11  (Obs.ID 00030379002), for a total exposure
time of 5.6  ks and a Suzaku \citep{mitsuda07} observation performed  on 2006 October 10
\citep[Obs.ID 401054010, see ][for further details]{morris09}.

 The  XRT   data  were   processed  with  standard   procedures  ({\sc
   xrtpipeline}  v.0.12.4), filtering  and  screening criteria,  using
 {\sc ftools}  in the  {\sc heasoft} package  (v 6.8). The  source was
 observed in Photon Counting mode \citep{hill04}.  We adopted standard
 grade filtering 0--12.  The  source events for spectral analysis were
 extracted from a circular region of 20 pixels radius (1 pixel=2.36'')
 centred on the source  position as determined with {\sc xrtcentroid}.
 The background  was extracted from  an annular region centred  on the
 source with radii  of 70 and 130 pixels,  respectively.  The spectrum
 was  analysed  using ancillary  response  files  generated with  {\sc
   xrtmkarf} and spectral redistribution matrix v011.

We  extracted Suzaku  scientific  data (Obs.ID  401054010) from  clean
events reprocessed  with the 2.0.6.13 standard  pipeline procedure. We
obtained four  spectra from the X-ray  Imaging Spectrometer \citep[XIS(0-4),][]{koyama07}.
Spectra  from the  front-illuminated CCD  (XIS0, XIS2  and  XIS3) were
combined    into    a    single    spectrum   (XIS023)    using    the
\textit{addascaspec} script,  while the XIS1 spectrum was  used in the
analysis as a different dataset.

 The  source  extraction region  was  centred  on  the CHANDRA  source
 position with a radius of 140 \arcsec. Background region was obtained
 from an  equal spatial  region adjacent to  the source  region.  Data
 from the  HXD/PIN instrument \citep{takahashi07}  were obtained using
 the \textit{hxdpinxbpi} script. The HXD/PIN background is the
   sum   of  a   time-variable   instrumental  background   (non-X-ray
   background,  NXB)  induced  by  cosmic  rays  and  trapped  charged
   particles  in   the  satellite  orbit  and   the  intrinsic  cosmic
   background (cosmic X-ray background, CXB).  We used the 2.0 version
   of  the  \textit{tuned} NXB  model  spectrum  released  by the  HXD
   instrument  team \citep[\texttt{LCFITDT}  model,][]{fukazawa09} and
   the CXB spectrum obtained from the \citet{boldt87} X-ray background
   model  convolved  with  the  PIN  response for  the  flat  emission
   distribution at the epoch of  this observation.  Given the lack of
 local  narrow features, spectra  from XIS  and HXD/PIN  were coarsely
 rebinned into  a spectrum of 84  (21) energy channels  for the XIS023
 (XIS1) and 9  energy channels for the HXD/PIN.   This choice leaves a
 S/N ratio  above 3 for each  energy channel. XIS  data are background
 dominated below 2  keV and above 10 keV, so  that these channels were
 not used in the spectral  analysis.  HXD/PIN spectrum was used in the
 12.0-40.0 keV  energy range.

Spectral fits  were performed using  {\sc xspec} v.12.6.0.   
Errors are at 90\,\% confidence level, if not stated otherwise.

%%%%%%%%%%%%%%%%%%%%%%%%%%%%%%%%%%%%%%%%%%%%%%%%%%%%%%%%%
\section{Spectral Analysis \label{spectral}}
%%%%%%%%%%%%%%%%%%%%%%%%%%%%%%%%%%%%%%%%%%%%%%%%%%%%%%%%%
BAT data  were collected  during the monitoring  campaign from
  2004  December to 2009  May, ISGRI  data from  2003 January  to 2009
  April, and  they both  track, within the  same temporal  window, the
  long-term  average  hard  X-ray  emission.   Swift-XRT  and  Suzaku
  observations,  as indicated  in  the previous  section, are  pointed
  observations taken  at different times. In order  to constrain the
broadband  spectrum we  distinguished, therefore,  the two  soft X-ray
observations performed  with XRT and Suzaku.  For \textit{Spectrum 1},
we hereafter  assume the  spectrum containing the  following datasets:
XRT, BAT,  ISGRI.  For \textit{Spectrum  2}, the spectrum  composed of
these datasets: Suzaku (XIS1, XIS023 and HXD/PIN), BAT and ISGRI.

We adopted  an inter-calibration multiplicative constant  to take into
account  flux   uncertainties  in   the  response  of   the  different
instruments.  A common  fit of the ISGRI and BAT  data does not reveal
any  systematic flux  mismatch.  Data could  satisfactorily be  fitted
using a power-law model and  the residuals did not show any systematic
difference in  the two datasets.   A calibration constant  between the
datasets (fixed  to one for the BAT  data and left free  for the ISGRI
data)  was  found to  be  consistent  with  one.  Therefore,  for  the
following analysis we fixed the inter-calibration factor between these
two instruments to one.

For  \textit{Spectrum 1},  the constant  was fixed  to 1  for  the XRT
dataset and left free for the BAT/ISGRI datasets ($C_{\rm bat}$).  For
\textit{Spectrum  2}, the  constant  was  fixed to  1  for the  XIS023
spectrum and  left free for  the HXD-PIN dataset ($C_{\rm  hxd}$), for
the  XIS1  dataset ($C_{\rm  xis1}$)  and  for  the BAT/ISGRI  dataset
($C_{\rm bat}$). The 1--10 keV absorbed/unabsorbed flux was
  (5.26$\pm$0.08/8.15$\pm$0.15)  $\times$  10$^{-11}$  erg  cm$^{-2}$
  s$^{-1}$ for the Swift-XRT observation, whereas it was (1.36$\pm$0.01/2.35$\pm$0.05)
  $\times$ 10$^{-11}$ erg cm$^{-2}$  s$^{-1}$ during the
  Suzaku  observation  (errorbars  take  also into  account the uncertainty
  associated with different modelizations of the spectra).

To model the data, we employed  the three most commonly used models to
fit   broadband  spectra   of  accreting   high-mass   X-ray  binaries
\citep[see][for     a      comprehensive     discussion     of     the
  models]{makishima99}:   a    cut-off   power   law    model   (model
\texttt{cutoffpl}),  a  power-law with  a  Fermi-Dirac cut-off  (model
\texttt{fdco}),   the   negative-positive   power-law   model   (model
\texttt{npex},  with  the positive  photon  index  fixed  to +2).  The
\texttt{cutoffpl}  model has three  main parameters:  the photon-index
($\Gamma$), the  cut-off energy (E$_{\rm cut}$),  the normalization of
the  power-law (N$_{\rm po}$);  the \texttt{fdco}  model has  one more
parameter, the  folding energy of  the power-law (E$_{\rm  fol}$); the
\texttt{npex} is  basically the  sum of two  power-laws with  a common
exponential cut-off.

We did not use the partial covering model  of a power-law spectrum
as  in   \citet{morris09},  because  this  model   is  not  physically
consistent with the nature of \src,  as the 6.732 d orbital period and
the 1069 s  spin period place unambiguously \src  in the wind-fed HMXB
zone  of the  P$_{\rm orb}$-P$_{\rm spin}$  diagram \citep[also  known as  the
  Corbet diagram,][]{corbet84}.

\subsection{\textit{Spectrum 1}}
In  the first column of Table\ref{tab1},  we present  the best-fitting  results  
adopting our set of spectral models for the \textit{Spectrum 1} datasets.

\begin{figure}
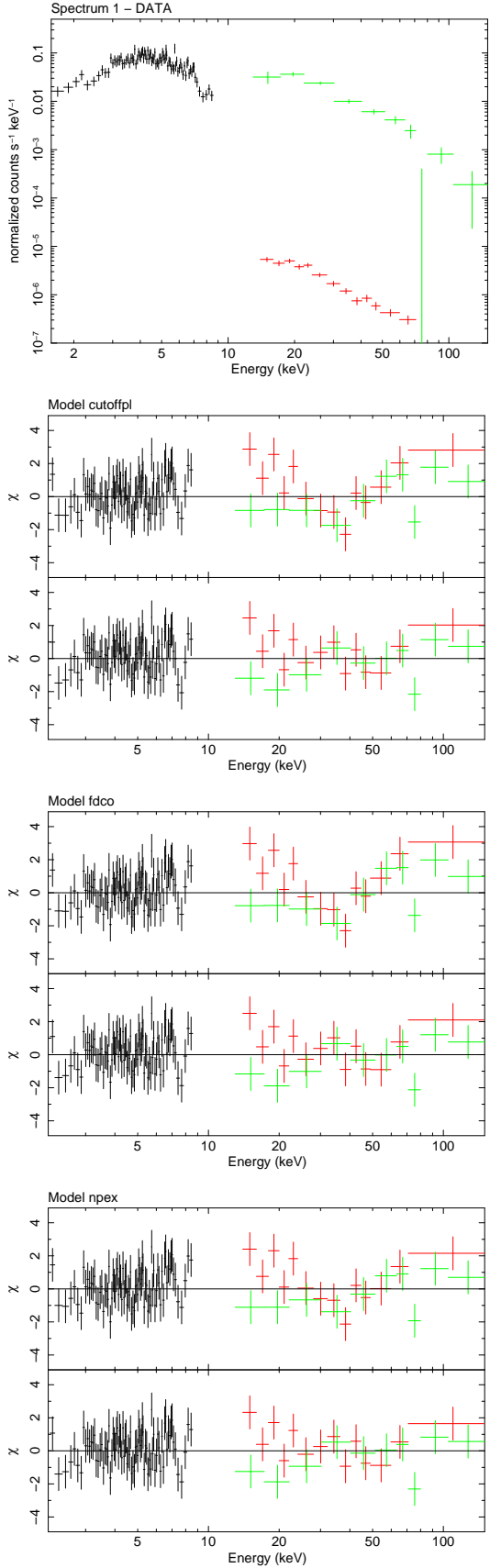

\begin{center}
\begin{tabular}{c}
\includegraphics[width=5.5cm,angle=-90]{data_spectrum1.ps} \\ 
\includegraphics[width=5.5cm,angle=-90]{spec1_cpl_res.ps} \\
\includegraphics[width=5.5cm,angle=-90]{spec1_fdco_res.ps} \\
\includegraphics[width=5.5cm,angle=-90]{spec1_npex_res.ps} 
\end{tabular}
\caption{\textbf{Upper panel:} datasets that compose the \textit{Spectrum
    1} analysis.  Black data:  XRT spectrum;  red  data: BAT
  spectrum;  green  data:   ISGRI  spectrum.   \textbf{Lower  panels}:
  residuals in  units of  $\sigma$ for the  continuum models  of Table
  \ref{tab1};  upper  mid-panels:  only continuum,  lower  mid-panels:
  continuum  with the addition  of a  cyclotron absorption  feature at
  $\sim$ 31 keV.}
\label{fig1}
\end{center}
\end{figure}

The  absorbed power-law with  a high  energy exponential  cut-off
gives  a  column density  \nh~  $=$  (7.0$\pm$1.2)
$\times$10$^{22}$  \cmmdue, $\Gamma=0.7\pm 0.3$  and a  cut-off energy
E$_{\rm   cut}=20\pm5$   keV.    This   model  already   proposed   in
\citet{hill08} is a better representation  of the data with respect to
the  more sophisticated  \texttt{fdco} model,  but we  found  that the
\texttt{npex} model provides  still a better fit to  the data, even if
the steepness  of the  power-law, the cut-off  energy and the  \nh are
consistent  in the  two models.   Irrespective of  the  model adopted,
broad   residuals  were   evident   between  30   and   40  keV   (see
Fig.\ref{fig1}).   Inspecting the  data, we  noted that  the residuals
were  mostly  driven by  the  residuals of  the BAT data, because of the 
better S/N ratio of these data in the hard X-ray band.
We have tentatively identified the shape of the residuals with a broad feature
in  absorption  and  we  repeated  the fits  introducing  a  cyclotron
absorption  feature  (\texttt{cyclabs}  component,  second  column  in
Table \ref{tab1}), which is expressed according to the formula:

\begin{equation}
CYCLABS(E)= D_{\rm c}\frac{(W_{\rm c}E/E_{\rm c})^2}{(E-E_{\rm c})^2+W_{\rm c}^2}
\end{equation}
\noindent
where $E_{\rm c}$, $D_{\rm c}$ and $W_{\rm c}$ are the cyclotron energy, depth and width
respectively \citep{makishima90}. We assume that the broad feature constitute
the fundamental and, given the low S/N at higher energies, we neglect higher harmonics.

The line parameters could not  be all adequately constrained. We found
that the width of the line strongly correlated with the cut-off energy
of  the  continuum  models,  resulting  in a  large  uncertainty.  We,
therefore, kept this parameter frozen to  the value of 10 keV, as this
was the most reasonable value that provided the lowest $\chi^2$ in all
the  models adopted.   This choice  will also  be  discussed \textit{a
  posteriori}.   The  energy  of   the  line,  for  the  \texttt{npex}
best-fitting model  is 33$\pm$4  keV, while the  depth of the  line is
0.5$\pm$0.2.  Similar  values were found  also adopting the  other two
models, indicating a weak correlation  with the broadband shape of the
continuum.  The addition of  this component significantly improved the
fit (reduced $\chi^2$ 1.16 against a value of 1.32 without the line).
To fit  CRSFs, a  Gaussian absorption model  (\texttt{gabs} in
  Xspec)  has been  often also  used \citep{coburn02}.   Replacing the
  \texttt{cyclabs} with the \texttt{gabs} model, did not significantly
  change  the conclusions  of our  analysis.  We  found again  a tight
  correlation between depth and $\sigma$ of the line. The best-fitting
  position of the line varied, according to the model adopted, between
  31 keV  and 35 keV. Keeping the  line $\sigma$ frozen at  5 keV, the
  best-fitting depth of  the line was found between  1 and 5. Relative
  errors on the best-fitting parameters  were of the same order of the
  corresponing  \texttt{cyclabs} analogues.   The $\chi^2$  values were
  found  marginally  worse  with   respect  to  the  models  with  the
  \texttt{cyclabs} component.

 The best-fitting  value of C$_{\rm  bat}$ is $\sim$0.2.   We verified
 that this  value is  consistent with the  intensity of the  BAT light
 curve during  the XRT observation,  a factor $\sim4$ higher  than the
 average source intensity along the 54 months of monitoring.

We show  in Fig.\ref{fig1} data  of \textit{Spectrum 1}  and residuals
for the models of first/second columns of Table\ref{tab1}.
\begin{table}
\caption{Spectral fitting results}
\label{tab1}
\begin{tabular}{lll|ll}
\hline
\hline
                           &  \multicolumn{2}{c}{Spectrum1} & \multicolumn{2}{c}{Spectrum2} \\
%\hline
                           &    \textit{No cyc} & \textit{+Cyc} & \textit{No cyc} & \textit{+Cyc} \\
                           &              \multicolumn{4}{c}{\texttt{cutoffpl}} \\
\hline
\nh                        & 7.0$_{-1.2}^{+1.5}$  & 5.5$_{-0.8}^{+1.5}$      & 7.7$_{-0.5}^{+0.4}$ & 7.5$\pm$0.4 \\
$\Gamma$                   & 0.7$\pm$0.3         &  0.4$_{-0.2}^{+0.4}$       & 1.1$\pm$0.1     & 1.13$\pm$0.12\\
N$_{\rm po}$  (10$^{-3}$)       & 5.4$_{-2.0}^{+4}$ & 2.7$_{-0.8}^{+2.5}$        & 2.5$\pm$0.5      & 2.4$\pm$0.4  \\
E$_{\rm cut}$ (keV)            &  20$\pm$5          & 19$_{-2}^{+6}$              & 24$_{-3}^{+4}$ & 27$_{-5}^{+6}$ \\
C$_{\rm bat}$                  &  0.23$_{-0.05}^{+0.08}$ & 0.18$_{-0.03}^{+0.07}$   & 1.18$\pm$0.14    &1.28$\pm$0.11 \\
C$_{\rm xis1}$ & & &1.09$\pm$0.03 &1.09$\pm$0.03 \\
C$_{\rm hxd}$  & & &0.92$\pm$0.13&0.92$\pm$0.15 \\

E$_{\rm cyc}$ (keV)            &                        & 32$\pm$4               &                  & 30$\pm$5\\
$D_{\rm c}$             &                               & 0.6$\pm$0.2            &                  & 0.45$\pm$0.17 \\
$\chi^2_{red}$ (d.o.f.) & 1.41 (106)                 & 1.18 (104)           &  1.24 (133)        & 1.12 (131) \\

                       &           \multicolumn{4}{c}{\texttt{fdco}} \\
\hline 
\nh                        & 7.3$_{-1.3}^{+1.5}$ & 6.0$_{-1.2}^{+1.1}$      & 8.0$\pm$0.6    &7.8$\pm$0.6    \\
$\Gamma$                   &0.9$\pm$0.3 & 0.6$\pm$0.3                    & 1.28$\pm$0.12   & 1.24$\pm$0.11   \\
N$_{\rm po}$  (10$^{-3}$)       &12$_{-5}^{+10}$  &   5.2$\pm$0.8              & 5.8$\pm$0.6     & 5.3$\pm$1.1 \\
E$_{\rm fol}$ (keV)            &18$_{-3}^{+5}$  &19$_{-3}^{+4}$                & 22$_{-2}^{+3}$   & 24$_{-3}^{+4}$  \\
E$_{\rm cut}$ (keV)            &$<$ 7.4  & $<$ 20                             &$<$8             &  $<$22   \\
C$_{\rm bat}$                  &0.22$_{-0.05}^{+0.08}$ &0.18$\pm$0.05          &1.13$\pm$0.11     & 1.24$_{-0.17}^{+0.22}$     \\
C$_{\rm xis1}$ & & &1.10$\pm$0.03 &1.10$\pm$0.03 \\
C$_{\rm hxd}$  & & &0.90$\pm$0.13 &0.91$\pm$0.15 \\
E$_{\rm cyc}$ (keV)            &       &32.4$\pm$3.5                         &                   &  31$\pm$5    \\
$D_{\rm c}$          & &0.6$\pm$0.2                                &                   & 0.50$\pm$0.17   \\
$\chi^2_{red}$ (d.o.f.)    &1.48 (105) & 1.21 (103)                      &1.33 (132)         & 1.17 (130)  \\
%\hline
                       &           \multicolumn{4}{c}{\texttt{npex}} \\
\hline
\nh                        & 7.3$_{-1.3}^{+1.1}$   & 5.8$_{-1.0}^{+1.2}$ & 7.7$\pm$0.6         & 7.5$\pm$0.5  \\
$\Gamma$                   & 0.81$_{-0.3}^{+0.15}$ & 0.45$\pm$0.3       & 1.05$_{-0.6}^{+0.1}$  & 1.01$\pm$0.16     \\
E$_{\rm cut}$ (keV)            & 15.3$_{-2.7}^{+0.7}$  & 16$_{-4}^{+6}$      & 16$_{-2}^{+4}$       & 17$_{-1}^{+6}$ \\
N$_p$\tablefootmark{a} (10$^{-8}$)          &6.4$_{-4.7}^{+1.7}$    & $<$70              & 1.3$_{-0.9}^{+0.2}$   & 1.0$_{-0.9}^{+0.5}$ \\
N$_n$\tablefootmark{b} (10$^{-3}$)          &6.1$_{-2.4}^{+2.8}$    & 3.1$\pm$0.1        & 2.4$\pm$0.5          & 2.2$\pm$0.4    \\
C$_{\rm bat}$                  & 0.28$_{-0.08}^{+0.06}$ & 0.20$\pm$0.06     & 1.35$\pm$0.14        & 1.41$\pm$0.18 \\
C$_{\rm xis1}$ & & 1.10$\pm$0.03 &1.10$\pm$0.03 \\
C$_{\rm hxd}$  & & 1.06$\pm$0.15 &1.04$\pm$0.16 \\
E$_{\rm cyc}$ (keV)            &                      & 33$\pm$4           &                      & 31$_{-12}^{+8}$\\
$D_{\rm c}$            &                     & 0.5$\pm$0.2         &                      & 0.30$_{-0.14}^{+0.19}$\\
$\chi^2_{red}$ (d.o.f.)    & 1.32 (105)            & 1.16 (103)         & 1.16 (132)           &  1.10 (130) \\
\hline
\hline
\end{tabular}
\tablefoot{
\tablefoottext{a}{Normalization of the power-law with positive index}
\tablefoottext{b}{Normalization of the power-law with negative index}
}

\end{table}

\subsection{\textit{Spectrum 2}}

We  used the  pointed Suzaku  observation to  perform  a complementary
analysis, checking  how a different  soft X-ray spectrum  could affect
the shape of  the hard X-ray spectrum.  The  analysis was performed in
analogy with  the steps described for \textit{Spectrum  1}.  The third
and fourth column of Table\ref{tab1}  show the results of the spectral
fitting for  the three  models, with and  without the addition  of the
cyclotron line.  The results are well in agreement with those obtained
for  \textit{Spectrum  1}.   Again,  the  best-fitting  model  of  the
continuum  emission is found  for the  \texttt{npex} model.   For this
model, the normalisation constant  C$_{\rm bat}$ is $\sim$1.4. A value
that is $\sim$6 times higher  with respect to \textit{Spectrum 1}; the
observed flux during the Suzaku observation is, in fact, $\sim$4 times
less with  respect to  the XRT observation.   Despite the drop  in the
soft X-ray flux no statistically significant variation in the spectral
shape of  the emission is  observed.  Best-fitting values for  the two
spectra  are in agreement,  although the  power-law steepness  is only
marginally  compatible,  indicating during  the  Suzaku observation  a
possible softer X-ray spectrum.

\begin{figure}
\begin{center}
\begin{tabular}{c}
\includegraphics[width=5.5cm,angle=-90]{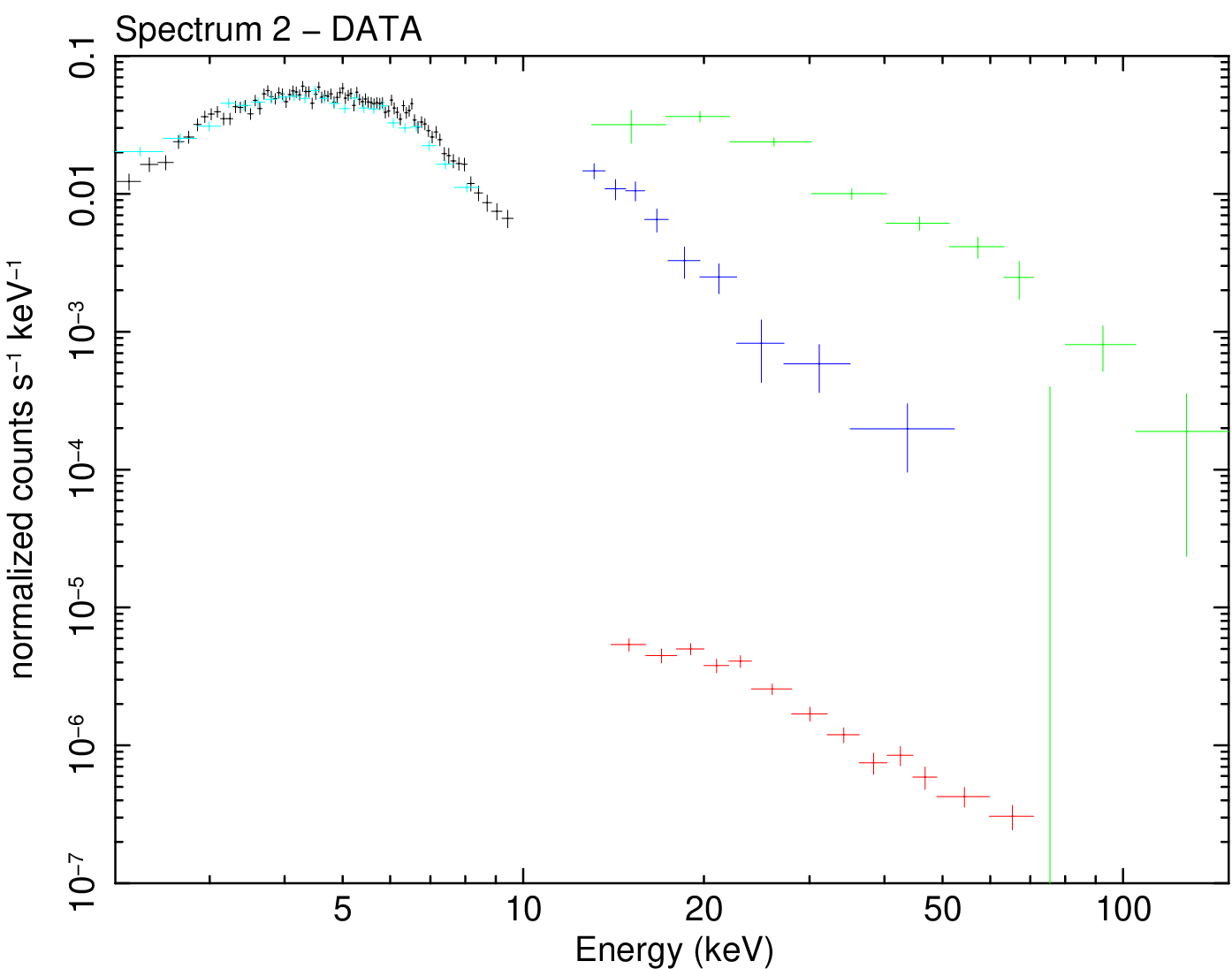} \\ 
\includegraphics[width=5.5cm,angle=-90]{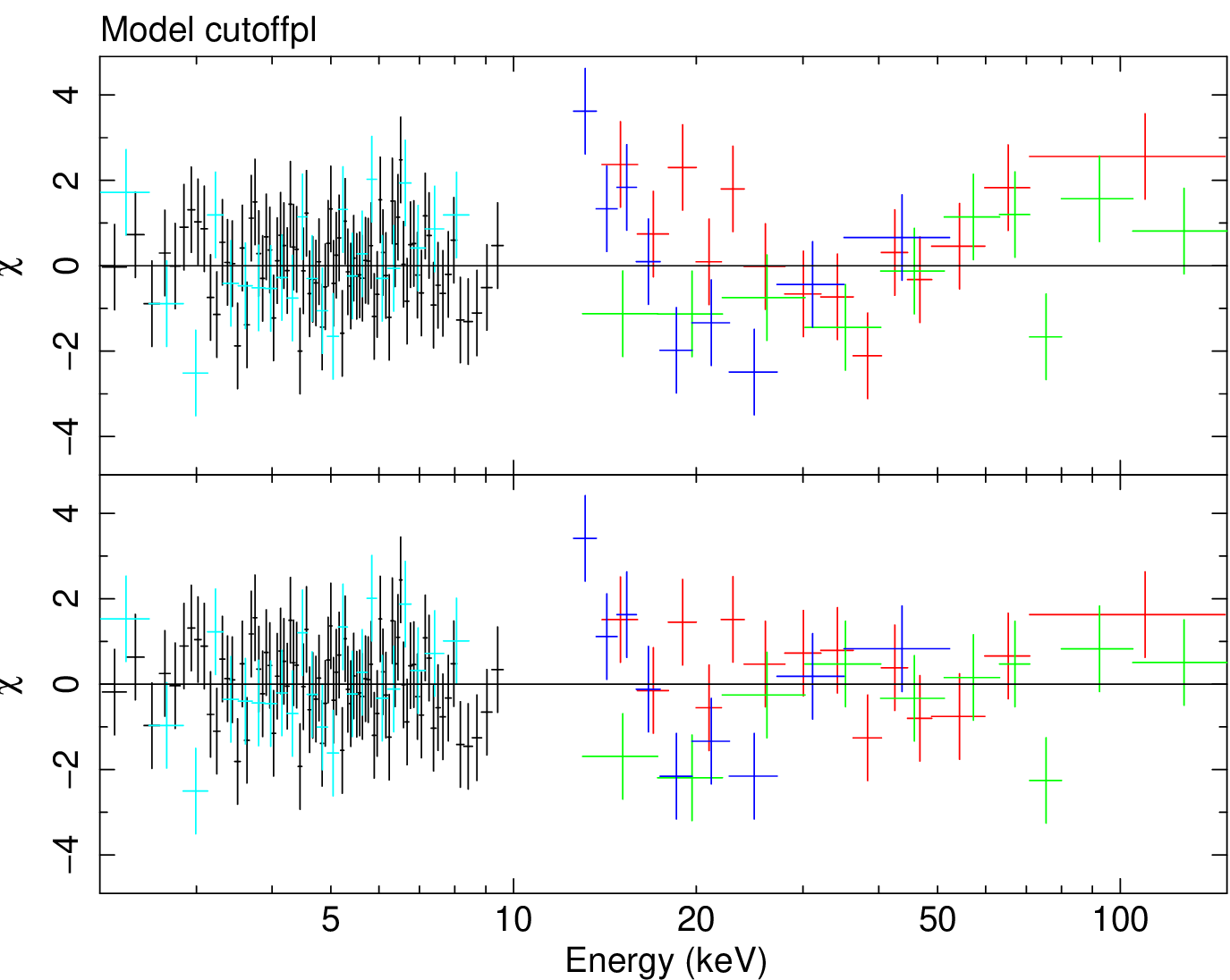} \\
\includegraphics[width=5.5cm,angle=-90]{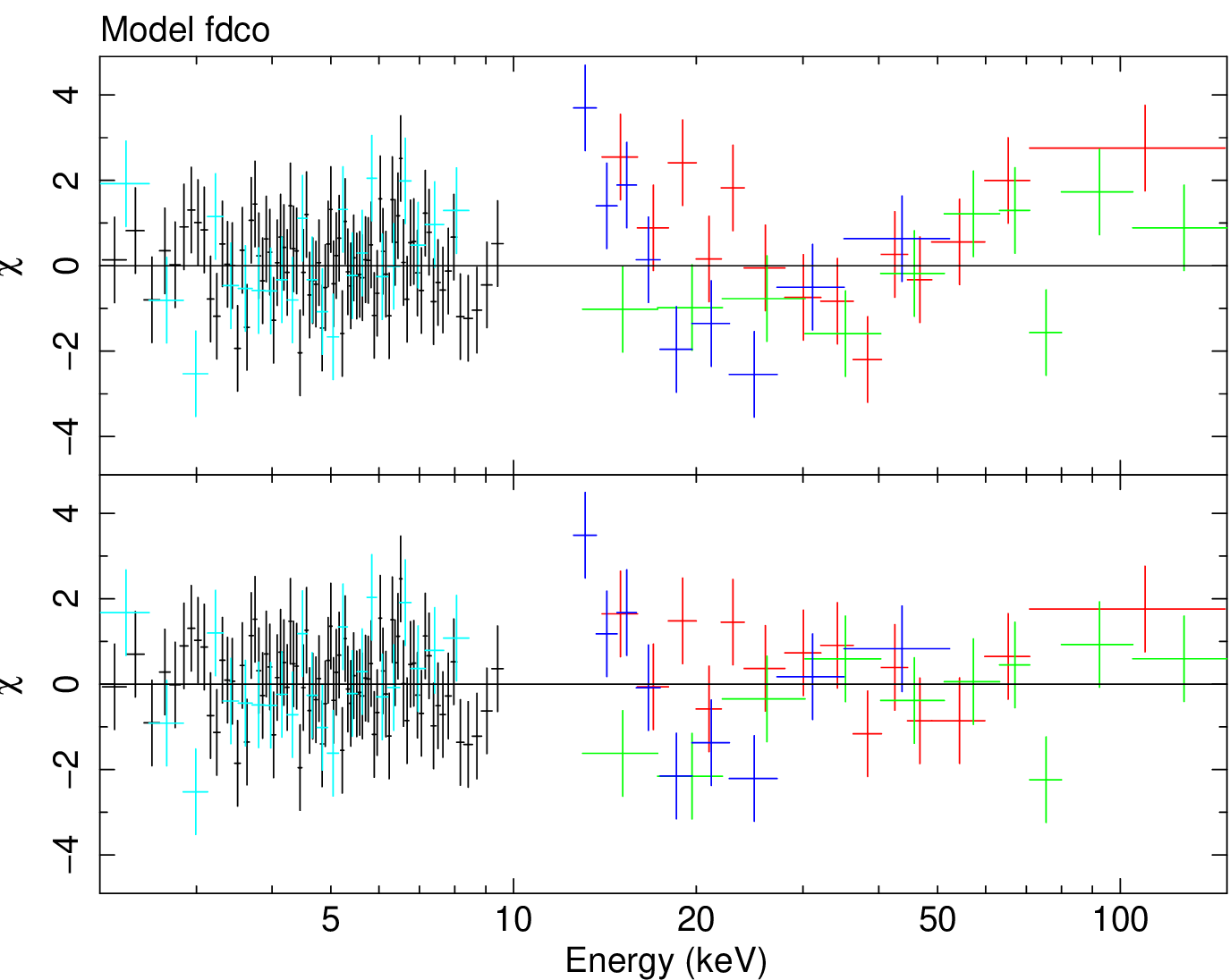} \\
\includegraphics[width=5.5cm,angle=-90]{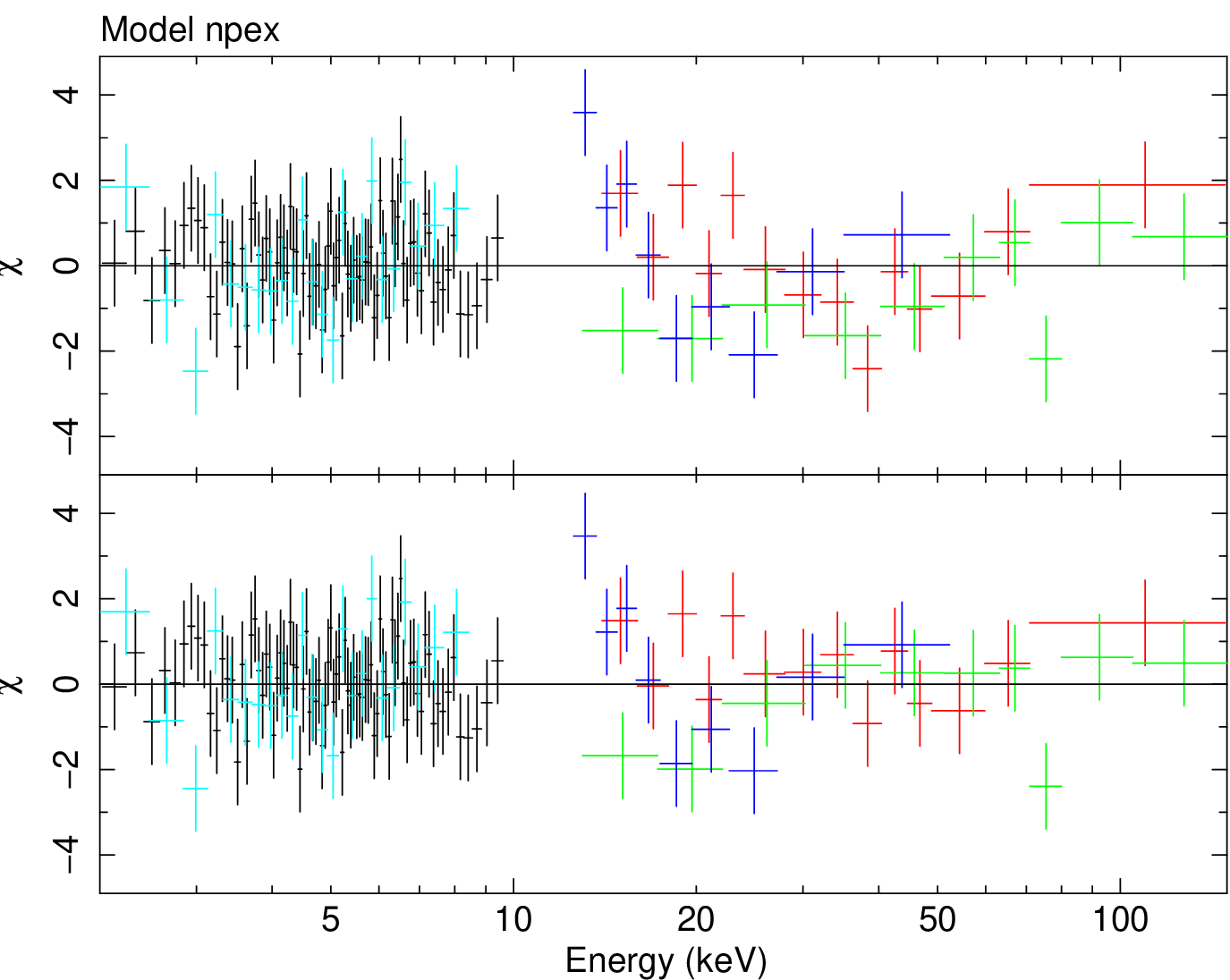} 
\end{tabular}
\caption{\textbf{Upper panel:} datasets that compose the \textit{Spectrum
    2} analysis.   Black  data:   XIS023
  spectrum; red  data: BAT spectrum;  green data: ISGRI
  spectrum;  blue  data: Suzaku/HXD-PIN spectrum; light  blue: XIS0  spectrum.
  \textbf{Lower panels}: residuals in units of $\sigma$ for
  the  continuum models  of Table  \ref{tab1}; upper  mid-panels: only
  continuum,  lower  mid-panels:  continuum  with the  addition  of  a
  cyclotron absorption feature at $\sim$ 31 keV.}
\label{fig2}
\end{center}
\end{figure}
The best  fit to the  continuum emission (\texttt{npex} model)  gave a
reduced  \chisq  =  1.16  (132  d.o.f.),  while  the  model  with  the
\textit{cyclabs} component gave a  reduced \chisq = 1.10 (130 d.o.f.).
The line  shape is found to  depend only marginally on  the soft X-ray
band.  The  cyclotron line  parameters are, however,  less constrained
with  respect  to \textit{Spectrum  1},  although best-fitting  values
remain well consistent.
We show in Fig.\ref{fig2}  data of \textit{Spectrum 2} and residuals for
the models of third and fourth column of  Table~\ref{tab1}.

\subsection{Calibration uncertainties of the BAT instrument}

Because the detection  of the cyclotron features is  mostly pivoted by
the  BAT high-energy  data,  we  checked for  possible  biases in  the
analysis due to incorrect background subtraction or systematics in the
BAT                                                            response
matrix\footnote{\url{http://swift.gsfc.nasa.gov/docs/swift/analysis/bat_digest.html}}
by comparing the extracted spectrum of \src with the spectra of nearby
X-ray sources having similar spectral shape.

We selected from  the 54-months BAT catalogue the  three HMXBs closest
to  \src: IGR J16393-4643,  IGR J16418-4532  and AX  J1700.2-4220.  We
extracted  the BAT long-term  spectra and  used pointed  Swift-XRT and
XMM-Newton  (Epic-PN spectrum  for  AX J1700.2-4220)  to obtain  their
broadband  X-ray spectra.   The  spectra of  IGR  J1639.1-4641 and  AX
J1700.2-4220  could be well  fitted using  a cut-off  power-law model,
while the  spectrum of IGR J16418-4532 required  a \texttt{npex} model
to be  satisfactorily fitted. In each  case the BAT data  did not show
any structure in the residuals in  the 30-40 keV range. Because of the
similarity  in  the  shape  of  the  continuum  emission,  the  common
background  estimate, and the  comparable statistics  in the  BAT data
among all  the sources of this  sample and \src, we  conclude that the
residuals shown in  the spectrum of \src cannot  be due to calibration
uncertainties of the  BAT instrument nor to an  incorrect estimate, or
subtraction, of the X-ray background.

\subsection{A method for testing the significance of the detection}

To estimate the statistical significance  of the residuals
present in the BAT data of \src, we adopted the following procedure (see also \citealt{suchy11}
for a similar approach) :
\begin{itemize}
\item  we  assumed  the  best-fitting  \texttt{npex}  model for \textit{Spectrum 1} in  Table
  \ref{tab1} without the cyclotron line as the \textit{null-hypothesis model};
\item we simulated accordingly a faked BAT spectrum having
the same S/N ratio of the real BAT spectrum;
\item  we  fitted again  the  datasets  of  \textit{Spectrum 1} (real  data  and
  simulated  BAT  spectrum) using  the  \texttt{npex}  model with  and
  without  the  addition  of   an  absorption  cyclotron  line  (width
  constrained  at 10  keV, with line  energy and
  optical depth as free parameters);
\item we calculated the difference in $\Delta \chi^2$ for
  the two best-fitting models: \texttt{npex} and \texttt{npex+cyclabs};
\item we repeated this procedure for other 10,000 faked BAT spectra;
\item we plotted with histograms  the $\Delta \chi^2$ difference versus the number
of spectra;
\item we repeated the same procedure for the \textit{Spectrum 2} datasets;
\end{itemize}

\begin{figure}
\begin{center}
\caption[]{Histograms  of the difference  in $\Delta  \chi^2$ obtained
  fitting  the datasets of  \textit{Spectrum 1}  and  \textit{Spectrum 2}, when the real BAT spectrum is substituted with
  a BAT simulated spectrum. The $\chi^2$ difference is calculated for the two
  best-fitting models: \textit{npex} and \textit{npex+cyc}.}
\label{histos}
\includegraphics[width=6cm,angle=-90]{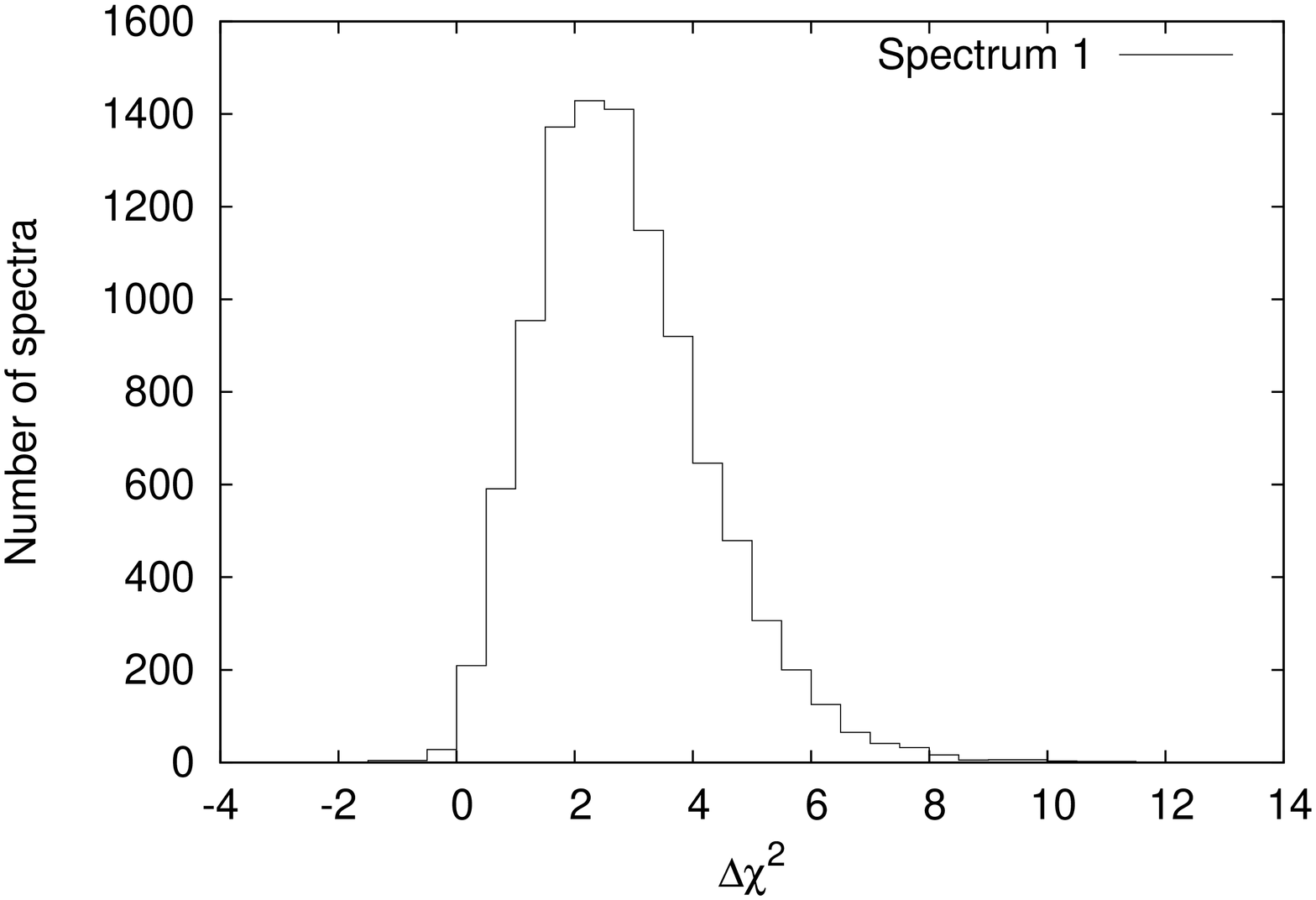} 
\includegraphics[width=6cm,angle=-90]{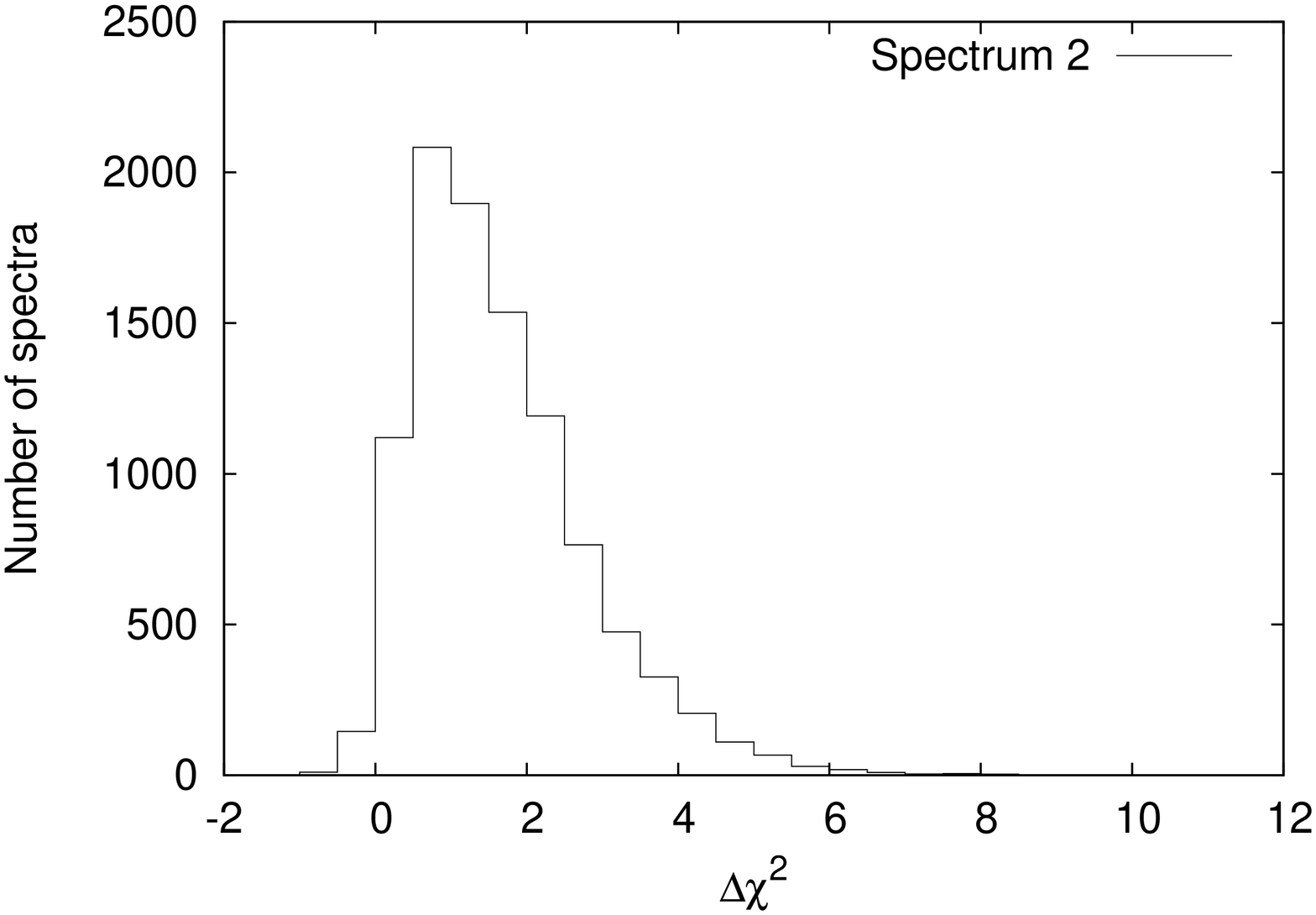} 
\end{center}
\end{figure}

In Fig.\ref{histos},  we show  the results of  our method.   In 10,000
spectra we did not obtain any difference in the $\Delta \chi^2$ as the
ones reported  in Table \ref{tab1},  both for the  \textit{Spectrum 1}
(observed  $\Delta   \chi^2$=19)  and  for   the  \textit{Spectrum  2}
(observed $\Delta \chi^2$=10) datasets.  The shape of the distribution
is  different  in the  two  cases: the  XRT  data  constrain less  the
broadband shape and  the addition of a broad  feature in absorption in
most cases improves the fit.  On the contrary the Suzaku data are less
sensible to  fitting improvements  even for the  addition of  a broad,
continuum-like,  feature.   This  is  expected since  the  statistical
weight  of the Suzaku  XIS ($\sim$  20,000 counts  in the  XIS spectra
versus the $\sim$ 2,000 counts of the XRT spectrum) data more strongly
pivot  the  overall  determination  of  the continuum  shape  and  the
$\chi^2$  result.   At  the  same  time, tighter  constraints  of  the
continuum shape  make random fluctuations as  artifacts less probable.
We note,  therefore, that  the detection level  as estimated  from the
difference in  the $\chi^2$  of the two  models (with and  without the
addition of  the cyclotron absorption feature)  in \textit{Spectrum 2}
real  data,  although in  absolute  value  less  than what  found  for
\textit{Spectrum 1}, is consistent and in agreement with the detection
level  obtained  from \textit{Spectrum  1}  data.   Moreover, we  have
estimated  this detection  level  for the  model (\texttt{npex})  that
provided the  lowest improvements for  the addition of  this features;
the other  spectral models  would have given  a much  higher detection
confidence. Because from the generation of 10,000 simulated spectra no
random fit improvement of the same  order, or greater, than the one we
obtained from real data was  obtained for both datasets, we estimate a
detection level  corresponding to  at least $\sim$3.9  gaussian sigmas
for the presence of a cyclotron line in our fits.

%%%%%%%%%%%%%%%%%%%%%%%%%%%%%%%%%%%%%%%%%%%%%%%%%%%%%%%%%
\section{Discussion \label{disco}}
%%%%%%%%%%%%%%%%%%%%%%%%%%%%%%%%%%%%%%%%%%%%%%%%%%%%%%%%%

There are 15 sources that clearly have shown CRSFs signatures in their
spectra. This number represents about  10\% of the total population of
HMXBs  detected   in  our  Galaxy  \citep{reig11}   and,  despite  the
considerable  amount of observing  time dedicated  by modern  and past
X-ray observatories, to this class of sources, new detections of CRSFs
have not considerably risen in the last decade.

We  have presented  new  results based  on  the analysis  of the  data
collected by  BAT during the first 54  months of the \sw  mission on a
persistent  sgHMXB  \src.   We  have  shown  that  the  long-term  BAT
collected  hard X-ray  spectra provides  an access  for  detecting the
presence of CRSFs in this kind of sources. This new approach relies on
a  series  of  assumptions   and  caveats  that  will  be  hereinafter
discussed.

The  broadband  spectrum, modelled  with  a positive-negative  cut-off
power-law model  provides the statistically  most favoured description
of the data in the examined datasets.  The residuals in the BAT energy
range show the presence of  an absorption feature between $\sim$28 and
$\sim$38 keV.   The feature is  not sensibly dependent on  the adopted
soft X-ray contribution, although we  have just exploited the only two
available to date pointed observations with XRT and Suzaku.  The broad
negative residuals are compatible with a resonant cyclotron absorption
feature with  an estimated significance  of the detection at  least at
$\sim$ 4 $\sigma$ level.

  Similar  features are  often  seen  in the  high  energy spectra  of
  sgHMXB.  They  are interpreted as cyclotron lines  produced near the
  magnetic poles of the  accretion-powered neutron star. These are due
  to resonant  scattering processes of  the X-rays by  electrons whose
  kinetic  energies are  quantised  in discrete  Landau energy  levels
  perpendicular to the B-field. Their detection is fundamental for the
  understanding of  the accretion mechanisms unto the  neutron star as
  it allows a direct measurement  of the magnetic field of the neutron
  star (or at  least a lower limit, depending on  the height where the
  resonant    line   absorption    is   effectively    produced,   see
  \citealt{nishimura11}).

We  consider here  that the  cyclotron absorption  takes place  at the
poles  of  the neutron  star,  taking  into  account the  relativistic
gravitational  redshift: $\rm  E_{\rm  cyc}^{\rm obs}$  = $\rm  E_{\rm
  cyc}(1 + \rm z)^{-1}$, with

\begin{equation}
(1 + z)^{-1} = \left( 1- \frac{2GM_{\rm X}}{R_{\rm X}c^2}\right)^{0.5}
\end{equation}

Assuming typical values for the NS  mass and radius (1.4 \msun and 10
km, respectively) and the observed  value of the cyclotron line energy
of 33$\pm$4  keV (\texttt{npex}  model for \textit{Spectrum  1}), this
implies a magnetic field of B$_{\rm NS}$ (3.7$\pm$0.4) $\times10^{12}$
Gauss at the surface of the NS.

The   energy  resolution  of   the  spectra   is  not   sufficient  to
independently constrain  line energy,  line width and  depth.  Because
the line energy and depth  could be relatively better constrained with
respect to the  line width, we choose to freeze this  value to 10 keV.
This value represents a  first-order guess assumed in consideration of
the  emerging relations  among  the parameters  that characterize  the
cyclotron line  shapes.  An empirical correlation  between line energy
and line width \citep{coburn02} is, in fact, observed, where cyclotron
lines detected between  30 keV and 40 keV  show preferably line widths
in  the  5--10 keV  range  (e.g.  in  MX  0656-072,  4U 0352+309,  XTE
J1946+274,  [see \citet{coburn02,mcbride06}]).   We tried  in  all the
fittings also to freeze the line width at 5 keV and 8 keV to check how
sensible was  the determination of this parameter.  We generally found
that the 10 keV guess always provided the lowest $\chi^2$, even if the
improvements  in  most  cases  were  marginal  (2$<\Delta  \chi^2<$5).
Another important empirical  correlation concerns the cut-off energies
of  the continuum  emission and  the  cyclotron line  energy, where  a
correlation is  observed with cut-off  energies typically at  half the
value of the cyclotron energy.   Also in this case, our results appear
in agreement with  the general trend that is  observed, with a cut-off
energy at  $\sim$15 keV and the  cyclotron energy at  $\sim$ two times
this value \citep{heindl04}.

Possible biases in this analysis are envisaged by the use of long-term
time  averaged spectra for  features that  may show  some variability.
Some luminosity-dependent  shifts of the cyclotron  line position were
in      fact      clearly      observed      in      some      sources
\citep{mihara04,staubert07,nakajima10}.      However     the     large
uncertainties in  the present analysis  estimates do not allow  to set
this  possibility into  investigation.  \src  has never  shown unusual
periods  of strong  changes  in its  luminosity,  nor any  outbursting
behavior as  in the  case of X0331+53,  where luminosity changed  by a
factor $\sim$  200 \citep{nakajima10}, so that we  argue that possible
shifts,  if   present,  should   be  within  our   quoted  error-bars.
Moreover, it is  to be noted that the  cyclotron feature at 45
  keV of  A0535+262 did  not show any  variations despite a  change in
  luminosity  of two orders  of magnitude  \citep{terada06}. Another
source of bias concerns the possibility of continuum spectral changes,
that, when averaged, could  result in spectral artifacts. The analysis
of  the  soft   X-ray  spectrum  from  the  XRT   and  Suzaku  pointed
observations are not sufficient to test any variability pattern in the
broadband  spectral properties;  use of  simplified, phenomenological,
models are  also possible source of  bias and a  more physically based
model should be employed to  also infer the physical properties of the
accretion  column   environment  \citep{becker07,schonherr07}  and  to
better constrain the shape of cyclotron resonance features.

\subsection{Conclusions}
We have presented  spectral analysis in the high-energy  X-ray band of
\src using long-term survey data from BAT and ISGRI, complemented with
pointed soft X-ray observations from XRT and Suzaku.  The BAT spectrum
is to date the highest S/N  spectrum of this source in the high-energy
X-ray band.  When the spectrum is complemented with soft X-ray data, a
broad absorption  feature is  detected, irrespective of  the broadband
model used to  fit the data.  We modeled the  feature assuming that it
is a  CRSF. This interpretations  is statistically acceptable  and the
physical interpretation  of the parameters  is in agreement  with what
expected from the whole sample of HMXBs that show similar features. We
estimated  the significance level  of the  detection $>$  3.9 $\sigma$
confidence level and next generation of future observatories dedicated
to the high X-ray band  (e.g.  NuSTAR and Astro-H) will likely provide
a more stringent confirm of its presence and a substantial improvement
in the determination of its shape.

%%%%%%%%%%%%%%%%%%%%%%%%%%%%%%%%%%%%%%%%%%%%%%%%%%%%%%%%%
\section*{Acknowledgements}
%%%%%%%%%%%%%%%%%%%%%%%%%%%%%%%%%%%%%%%%%%%%%%%%%%%%%%%%%
%
The authors acknowledge financial contribution from the agreement ASI-INAF I/009/10/0 

\bibliographystyle{aa}
\bibliography{refs}

\end{document}